\documentclass[aps,prl,superscriptaddress, floatfix, twocolumn]{revtex4-1}

\usepackage{amsmath}
\usepackage{amssymb}
\usepackage{wasysym}
\usepackage{hyperref}
\usepackage{graphicx}
\usepackage{color}
\usepackage{physics}
\usepackage{siunitx}
\usepackage{mathrsfs}
\usepackage[english,nomargin,inline,marginclue,draft]{fixme}
\fxusetheme{colorsig}
\FXRegisterAuthor{cg}{acg}{CG}  
\makeatletter
\renewcommand*\FXLayoutInline[3]{%
  {\@fxuseface{inline}\ignorespaces{\color{fx#1}[#3: #2]}}}
\makeatother

\renewcommand{\l}{\left(}
\renewcommand{\r}{\right)}

\begin{document}

\title{Imaging magnetic polarons in the doped Fermi-Hubbard model}

\author{Joannis~Koepsell}
\email{joannis.koepsell@mpq.mpg.de}
\author{Jayadev~Vijayan}
\author{Pimonpan~Sompet}
\affiliation{Max-Planck-Institut f\"{u}r Quantenoptik, 85748 Garching, Germany}
\author{Fabian~Grusdt}
\affiliation{Department of Physics, Harvard University, Cambridge, Massachusetts 02138, USA}
\affiliation{Department of Physics, Technical University of Munich, 85748 Garching, Germany}
\author{Timon~A.~Hilker}
\thanks{present address: Cavendish Laboratory, University of Cambridge, JJ Thomson Avenue, Cambridge CB3 0HE, United Kingdom}
\affiliation{Max-Planck-Institut f\"{u}r Quantenoptik, 85748 Garching, Germany}
\altaffiliation{test}
\author{Eugene~Demler}
\affiliation{Department of Physics, Harvard University, Cambridge, Massachusetts 02138, USA}
\author{Guillaume~Salomon}
\affiliation{Max-Planck-Institut f\"{u}r Quantenoptik, 85748 Garching, Germany}
\author{Immanuel~Bloch}
\affiliation{Max-Planck-Institut f\"{u}r Quantenoptik, 85748 Garching, Germany}
\affiliation{Fakult\"{a}t f\"{u}r Physik, Ludwig-Maximilians-Universit\"{a}t, 80799 M\"{u}nchen, Germany}
\author{Christian~Gross}
\affiliation{Max-Planck-Institut f\"{u}r Quantenoptik, 85748 Garching, Germany}


\date{\today}

\begin{abstract}
Polarons are among the most fundamental quasiparticles emerging in interacting many-body systems, forming already at the level of a single mobile dopant \cite{Alexandrov2010}. In
the context of the two-dimensional Fermi-Hubbard model, such polarons are predicted to form around charged dopants in an antiferromagnetic background in the low doping regime close to the Mott insulating state \cite{SchmittRink1988, Shraiman1988, Sachdev1989, Kane1989, Dagotto1989, Martinez1991, Grusdt2018}. Macroscopic transport and spectroscopy measurements related to high $T_{c}$ materials have yielded strong evidence for the existence of such quasiparticles in these systems \cite{Zhou2006, Orenstein2007}. Here we report the first microscopic observation of magnetic polarons in a doped Fermi-Hubbard system, harnessing the full single-site spin and density resolution of our ultracold-atom quantum simulator. We reveal the dressing of mobile doublons by a local reduction and even sign reversal of magnetic correlations, originating from the competition between kinetic and magnetic energy in the system. The experimentally observed polaron signatures are found to be consistent with an effective string model at finite temperature \cite{Grusdt2018}. We demonstrate that delocalization of the doublon is a necessary condition for polaron formation by contrasting this mobile setting to a scenario where the doublon is pinned to a
lattice site. Our work paves the way towards probing interactions between polarons, which may lead to stripe formation, as well as microscopically exploring the fate of polarons in the pseudogap and bad metal phase.
\end{abstract}

\maketitle

\section*{Introduction}
\begin{figure}
\centering
\includegraphics{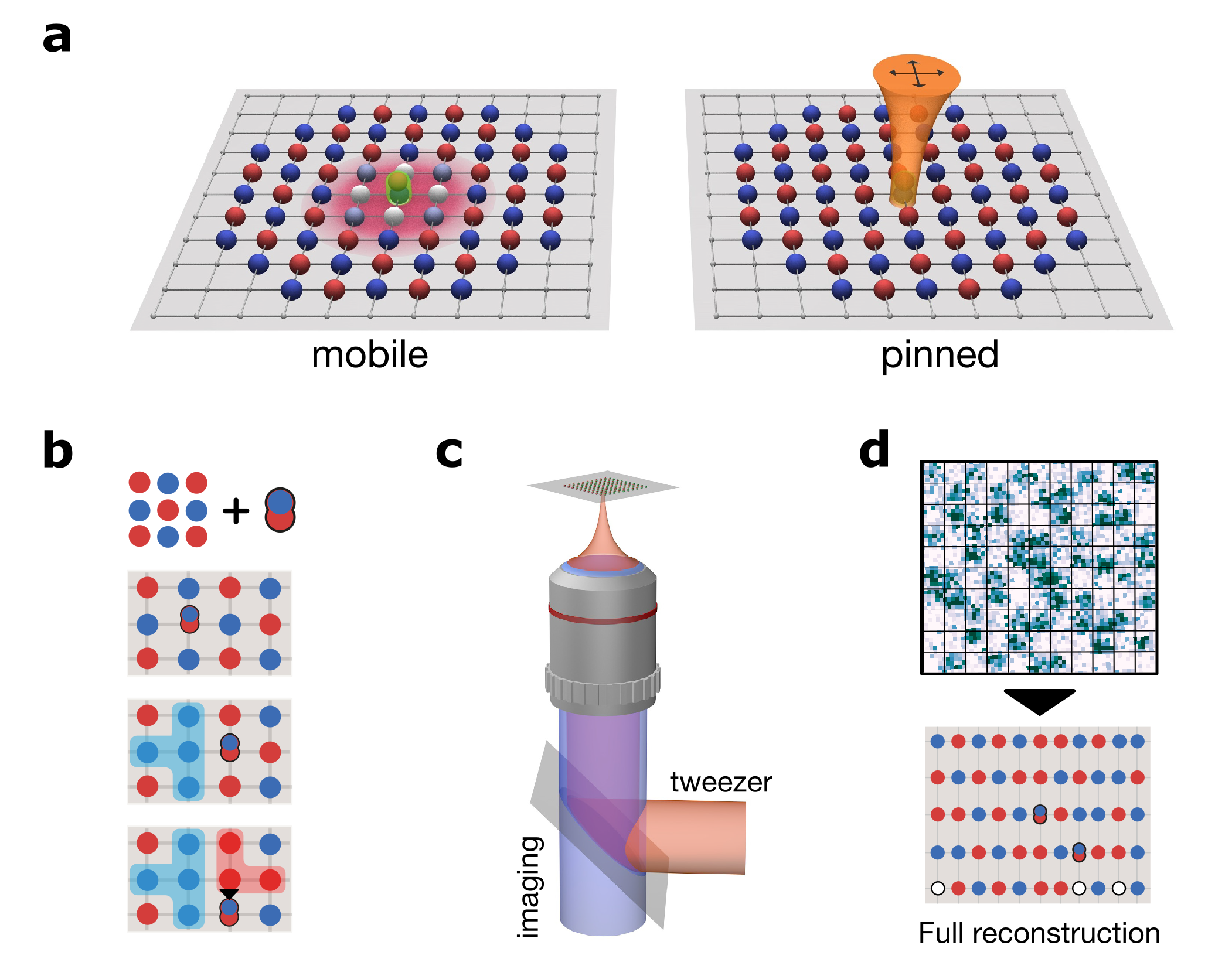}
\caption{{\bf Mobile and immobile dopants with ultracold atoms.} 
\textbf{a}, We experimentally study two-dimensional Fermi-Hubbard systems with mobile (left) or immobile (right) doublons, using a quantum gas microscope.
\textbf{b}, The delocalization of a mobile doublon in the antiferromagnetic background of the Fermi-Hubbard model around half-filling leads to a distorted spin order. With increasing amount of delocalization, antiferromagnetically aligned spin pairs are turned into ferromagnetic ones (see shading). As a consequence of this competition, theoretical and experimental evidence points towards polaron formation (see text).
\textbf{c}, To create immobile localized doublons, we overlap an attractive laser beam (orange) with the imaging light (blue) and focus it down to a single lattice site.
\textbf{d}, Each captured image corresponds to a projected many-body quantum state. Employing our local Stern-Gerlach technique, we fully resolve spin and density. This allows for the local investigation of the spin environment around doublons. As indicated in the reconstruced image, the Fermi-Hubbard model is implemented with anisotropic lattice spacings $a_{y}=2\cdot a_{x}$, but equal tunneling amplitudes $t_{y}=t_{x}$.
\label{fig:fig1}}
\end{figure}

An electronic charge carrier dressed by a local polarization of the background environment is called a polaron. Such quasiparticles usually occur in materials with a strong coupling between mobile charge carriers and collective modes of the background, such as phonons, magnons or spinons \cite{Alexandrov2010}. These materials furthermore often possess exotic properties, such as spin currents in organic semiconductors \cite{Watanabe2014}, colossal magnetoresistance in the manganites \cite{Ramirez1997}, pseudogaps in transition metal oxides or high $T_{c}$ superconductivity in the cuprates \cite{Lee2006}. Even though these phenomena still pose many open questions regarding their microscopic description, some of them can be attributed to polarons while others can emerge as a result of multiple interacting polarons \cite{Trugman1988, Schrieffer1989, Teresa1997, Verdi2017}. The most prominent and conceptually simple electronic model for high $T_c$ cuprates is the two-dimensional doped Fermi-Hubbard model, in which an interplay between the kinetic energy of doped charge carriers and a magnetic background supports the formation of magnetic polarons at the single dopant level. The model consists of spin-$\frac{1}{2}$ fermions hopping on a two-dimensional lattice with nearest-neighbour hopping amplitude $t$ and on-site repulsion $U$ between opposite spins. At half filling it reduces to a Mott insulating state with antiferromagnetic (AFM) correlations, due to an effective superexchange spin coupling $J=4t^2/U$. Upon hole or particle doping, the dopants can lower their kinetic energy by delocalization, which reduces the background AFM correlations (see Fig. \ref{fig:fig1}). As a consequence of the growing magnetic cost with increasing delocalization, theoretical simulations of single dopants in the related $t$-$J$ model predict the formation of a magnetic polaron \cite{Kane1989, SchmittRink1988, Dagotto1989, Grusdt2018, Martinez1991, Shraiman1988, Sachdev1989}, in which the dopant surrounds itself with a local cloud of reduced AFM correlations (see Fig.\ref{fig:fig1}a). Since this cloud can move only on a timescale of order $1/J$, the polaron acquires an increased effective mass as $J < t$. Spectroscopic measurements of undoped cuprates have experimentally probed this single dopant regime. Even though measurements of bandwidths or quasiparticle weights are compatible with the formation of polarons \cite{Zhou2006, Orenstein2007}, a direct microscopic image of such dressed mobile charge carriers at the single particle level is still lacking. Furthermore, the evolution from individual polarons into the pseudogap or strange metal phase upon higher doping concentration is still subject to controversy, allowing for a rich diversity of theoretical approaches \cite{Anderson1987, Kivelson1987, Auerbach1991, Punk2015}.

Quantum gas microscopy has enabled the direct, model-free real-space characterization of strongly correlated quantum many body systems. In cold-atom lattice simulators \cite{Gross2017}, this technique has proven its potential to shed new light on the Fermi-Hubbard model, including the detection of long-ranged spin correlations \cite{Mazurenko2017}, charge and spin transport \cite{Brown2018, Nichols2018} in two dimensions as well as incommensurate magnetism \cite{Salomon2018} in one dimension.
Employing the full spin and density resolution of our setup \cite{Boll2016}, we experimentally confirm the presence and microscopic structure of magnetic polarons in the small doping regime of our atomic Fermi-Hubbard system. We observe how mobile doublons are surrounded by a local distortion of antiferromagnetic correlations. Exact diagonalization of the $t$-$J$ model as well as an effective string model \cite{Grusdt2018} predict similar qualitative features of the magnetic polaron as measured in the experiment.  In contrast, by confining a doublon to a single lattice site with an optical tweezer, we find that immobile doublons lead to a local enhancement of AFM spin correlations. 

\begin{figure*}
\centering
\includegraphics{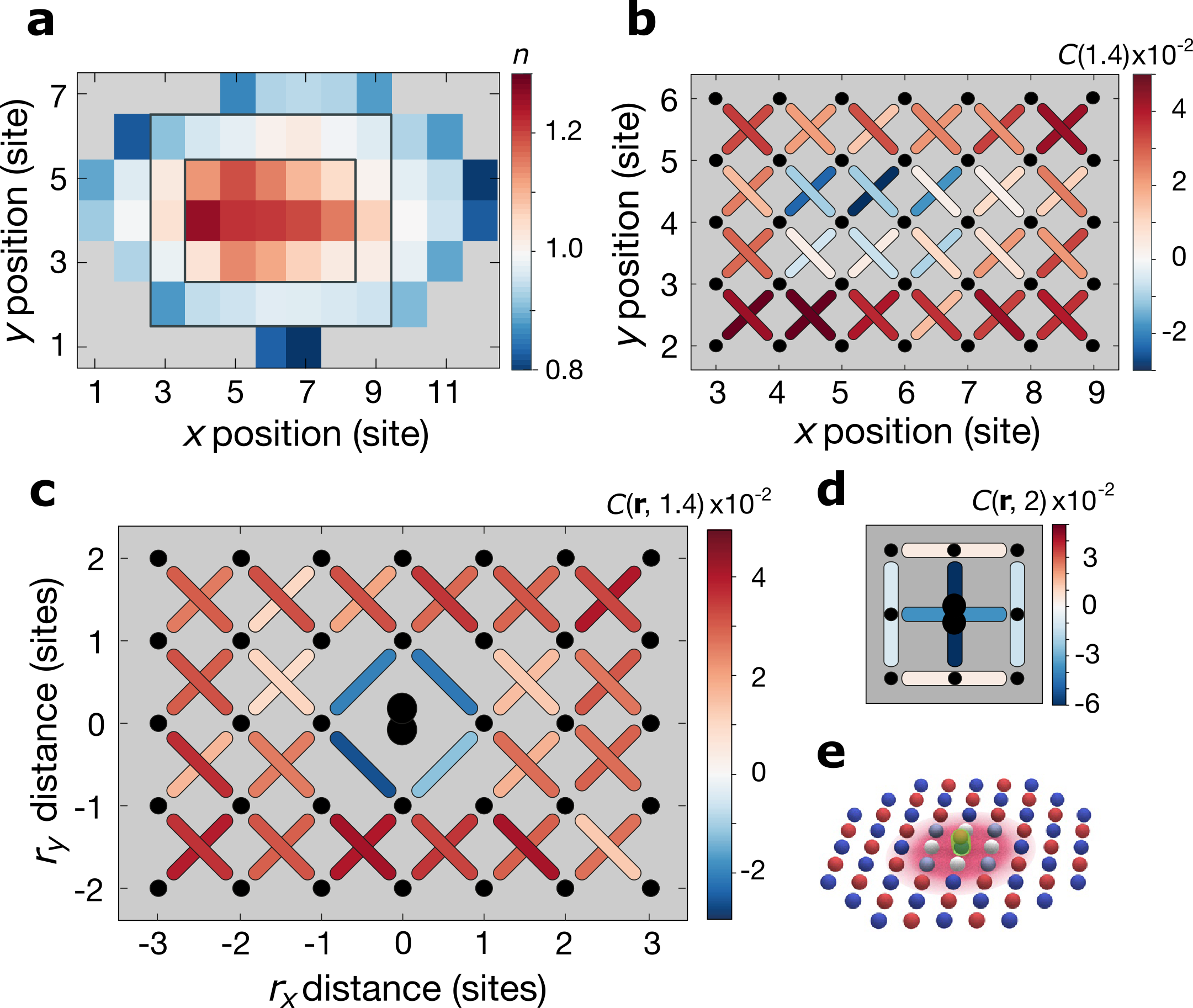}
\caption{\textbf{Mobile doublons dressed by local spin disturbance. }
\textbf{a}, Density distribution for the scenario of mobile doublons. In the doped region (inner black box) on average two doublons delocalize on an area of $5 \times 3$ sites.
\textbf{b}, Diagonal spin correlations in the central region of interest (outer black box in \textbf{a}), represented by bonds connecting the two sites (black dots) to be correlated. In the center a clear reduction of correlations from the  positive antiferromagnetic background value is visible. In the area of highest doublon density correlations even flip sign and become negative.
\textbf{c}, Spin correlations sorted by distance from doublons reveal the formation of magnetic polarons. In the reference frame of mobile doublons (indicated in the center by a double black circle) diagonal correlations become negative only in their immediate vicinity.
\textbf{d}, Next-nearest-neighbour spin correlations across and next to mobile doublons. Similar to the case of diagonal correlations (c), the correlations across doublons are sign flipped w.r.t the antiferromagnetic background value.
\textbf{e}, Our experimental results confirm the picture of a polaron, in which doublons are dressed by a local spin distortion.
\label{fig:fig2}}
\end{figure*}

\section*{System preparation and correlation functions}
In our experiment we prepared a balanced mixture of the two lowest hyperfine states of $^{6}\text{Li}$ and adiabatically loaded around 70 atoms into an anisotropic two dimensional square lattice with spacings $(a_{x},a_{y})=(1.15,\, 2.3)\,\mu m$ and depths ($8.6\,E_{r}^{x}, 3\,E_{r}^{y}$), where $E_{r}^{i}=h^{2}/8ma_{i}^{2}$ is the recoil energy of the respective lattice. The system is well described by the two-dimensional Fermi-Hubbard model with approximately equal tunnelling amplitudes in both directions of  $t_{y}/h \simeq t_{x}/h=170 \,\text{Hz}$ (see supplement). We tuned the interaction $U$ by using the broad Feshbach resonance in $^{6}\text{Li}$, such that $U/t_{i}=14(1)$, leading to a superexchange coupling $J/h=50(5)\,\text{Hz}$. We estimate the temperature of the system to be $T \simeq 1.4\,J$ by comparison of spin correlations in the half filled case with numerical linked-cluster expansion calculations of Khatami \textit{et al}. \cite{Khatami2011} (see supplement). In our study, we separately realized the setting of mobile and immobile dopants (see Fig. \ref{fig:fig1}) by doping with doublons. This has an experimental advantage compared to hole doping, because doublons are trapped by our confining potential and avoids contamination of the signal by false holes created during the detection. Mobile doublons were prepared by an increased chemical potential, resulting in delocalized doublons in the center of our harmonically confined lattice with trapping frequency of about $\omega/(2\pi)=250\,$Hz. For the preparation of immobile doublons we used a tightly focused laser beam (tweezer) at $702\,$nm with a waist of about $0.5\, \mu$m to form an attractive deep local potential. By shining the tweezer onto a single lattice site, the deep potential leads to an artificially created trapped doublon at that site (see Fig. \ref{fig:fig1}a). Our detection method \cite{Boll2016} allows us to simultaneously reconstruct the local spin and density within a single snapshot (see Fig. \ref{fig:fig1}d). In this way, we can separate the spin and density sector by measuring local spin correlations between two sites at positions $\textbf{r}_1$ and $\textbf{r}_2$ that are singly occupied (indicated by the filled circles below)
\begin{equation}
 C(\textbf{r}_1,\textbf{r}_2 )=4 \langle S^z_{\textbf{r}_1} S^z_{\textbf{r}_2}  \rangle_{  \scalebox{0.65}{\newmoon}_{\textbf{r}_1}\scalebox{0.65}{\newmoon}_{\textbf{r}_2}}.
 \label{eq:spincorr}
\end{equation}
We define the value of $C(\textbf{r}_{1},\textbf{r}_{2})$ as the \textit{bond strength} between $\textbf{r}_1$ and $\textbf{r}_2$. The spin environment around doublons can be investigated with a three-point doublon-spin correlator, which measures the two-point spin correlation as a function of a detected doublon at a third position $\textbf{r}_0$
\begin{align}
C(\textbf{r}_0;\textbf{r}_1,\textbf{r}_2 )&=4 \langle S^z_{\textbf{r}_1} S^z_{\textbf{r}_2}  \rangle_{ \overset{\scalebox{0.65}{\newmoon}}{\scalebox{0.65}{\newmoon}}  _{\textbf{r}_0}  \scalebox{0.65}{\newmoon}_{\textbf{r}_1}\scalebox{0.65}{\newmoon}_{\textbf{r}_2}} \nonumber \\
\equiv C(\textbf{r}_0;\textbf{r},\textbf{d} )&=4 \langle S^z_{\textbf{r}_0+\textbf{r}-\frac{\textbf{d}}{2}} S^z_{\textbf{r}_0+\textbf{r}+\frac{\textbf{d}}{2}}  \rangle_{ \overset{\scalebox{0.65}{\newmoon}}{\scalebox{0.65}{\newmoon}}  _{\textbf{r}_0}  \scalebox{0.65}{\newmoon}_{\textbf{r}_0+\textbf{r}-\frac{\textbf{d}}{2}}\scalebox{0.65}{\newmoon}_{\textbf{r}_0+\textbf{r}+\frac{\textbf{d}}{2}}}.
\label{eq:polaroncorr}
\end{align}
Here, the correlator is expressed in terms of the \textit{bond length} $\textbf{d}=\textbf{r}_2-\textbf{r}_1$ of the spin correlation and \textit{bond distance} $\textbf{r}=(\textbf{r}_1+\textbf{r}_2)/2-\textbf{r}_0$ from the doublon. This three-point correlator can be pictorially understood as  defining the origin in each snapshot as the position of a detected doublon and calculating arbitrary two-point spin correlations as a function of distance from that point. For a magnetic polaron, this correlator is expected to reveal the strongly altered spin correlations in the immediate vicinity of doublons (i.e. for small bond distances $\textbf{r}$). The remainder of this article will focus on the analysis of $C(\textbf{r}_0;\textbf{r},\textbf{d})$ for nearest-neighbour $|\textbf{d}|=1$ (NN), diagonal $|\textbf{d}|=1.4$ and next-nearest-neighbour $|\textbf{d}|=2$ (NNN) spin correlations as a function of bond distance $\textbf{r}$ from doublons. To minimize the contribution of doublon-hole fluctuations to our signal, we neglect double occupations with holes as nearest-neighbours (see supplement).

\section*{Polaron}
To study the doped system, we set the chemical potential, such that a doped region of $5 \times 3$ sites forms with on average 1.95(1) doublons present per experimental realization (see Fig.\ref{fig:fig2}a). First, we consider the effect of doping on standard two-point diagonal spin correlations (Eq. \ref{eq:spincorr}). We find that in the doped region AFM correlations are strongly reduced (see Fig. \ref{fig:fig2}b) and diagonal correlations may even switch their sign \cite{Parsons2016,Cheuk2016}. To reveal the presence of polarons as the microscopic origin of this effect, we analyze the correlator defined in Eq. \ref{eq:polaroncorr} to evaluate spin correlations as a function of bond distance $\textbf{r}$ from doublons. For each experimental snapshot, doublons are detected at different positions $\textbf{r}_0$. We average the correlator of Eq. \ref{eq:polaroncorr} over all positions in the doped region and obtain the average spin correlation around a single doublon $C(\textbf{r},\textbf{d})$ as displayed in Fig. \ref{fig:fig2}c for diagonal correlations ($|\textbf{d}|=1.4$). Remarkably, we observe the dressing of doublons with a spin disturbance, which is much stronger than visible on standard dopant unresolved two-point spin correlations. This confirms the picture of a magnetic polaron and therefore two-point spin correlations in the doped region in Fig. \ref{fig:fig2}b are a result of an occasional local distortion through a polaron (see supplement) and not a global reduction of AFM correlations. The strong effect on the magnetic correlations is even more pronounced in NNN correlations ($|\textbf{d}|=2$) across doublons, which reverse their sign with an amplitude a factor of two larger compared to diagonal ones (see Fig. \ref{fig:fig2}d). We attribute the stronger signal to the shorter minimal bond distance from doublons for the NNN correlation  $|\textbf{r}|=0$, compared to the minimal bond distance of $|\textbf{r}|=\frac{1}{\sqrt{2}}$ for the diagonal correlation (for NN correlations see supplement).

\begin{figure}
\centering
\includegraphics{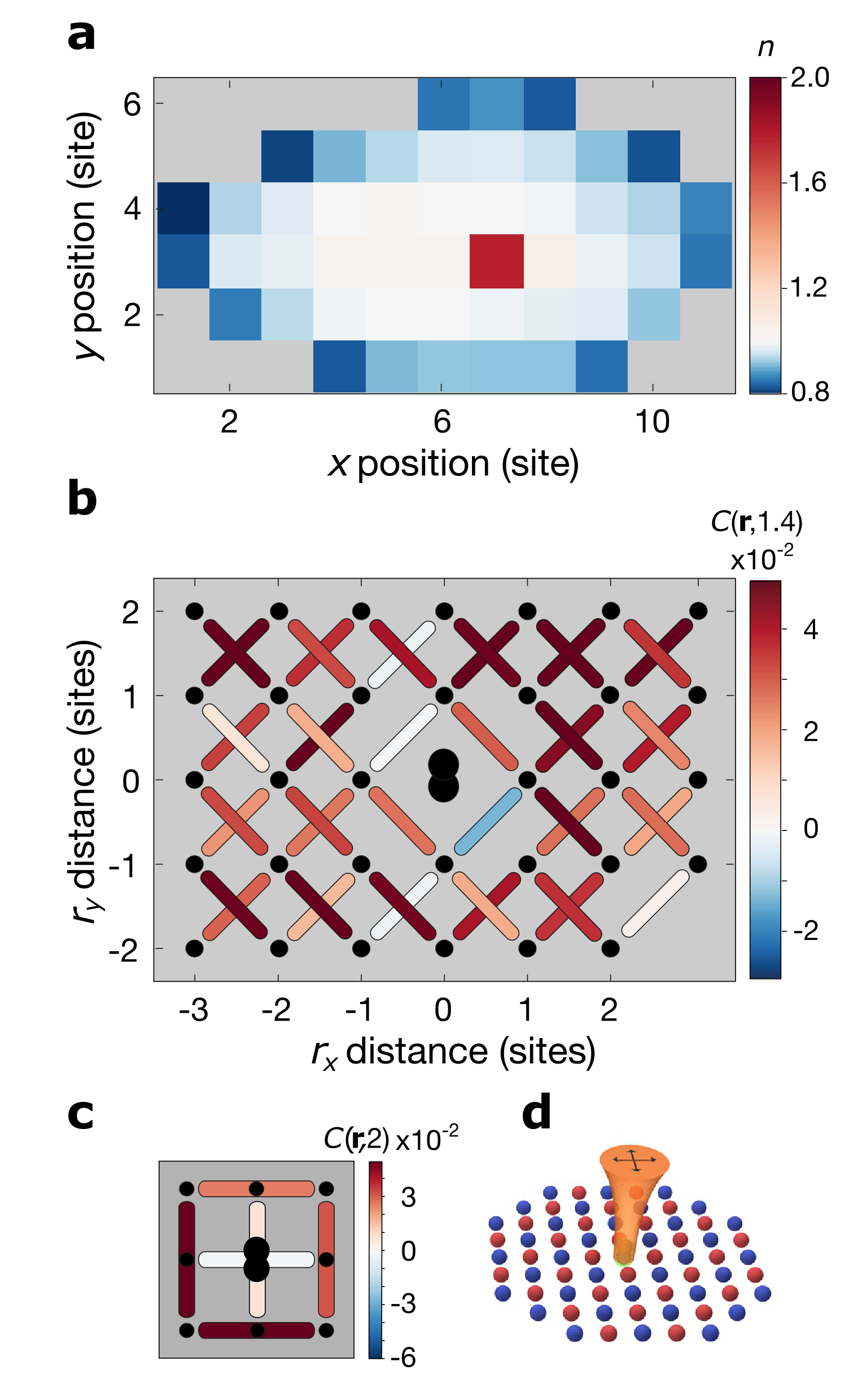}
\caption{\textbf{Spin correlations around trapped doublons. }
\textbf{a}, Density distribution for the scenario of immobile doublons. An attractive laser beam (tweezer) at $702\,$nm, focused to a single site, increases the density in an undoped system at a specific site artificially to about 1.77(1)
\textbf{b}, Diagonal spin correlations around doublons trapped in the tweezer. The sign flipped spin distortion vanishes compared to the mobile case. 
\textbf{c}, Next-nearest-neighbour spin correlations across and next to immobile doublons. While spins across the trapped doublon are uncorrelated, correlations neighbouring the trapped doublon are slightly enhanced, compared to the background value (see Fig.\ref{fig:fig4}c).
\textbf{d}, Trapping doublons with a tweezer beam stops the competition between kinetic and magnetic energy and prevents polaron formation.
}
\label{fig:fig3}
\end{figure}

\begin{figure}
\centering
\includegraphics[width=8.9cm]{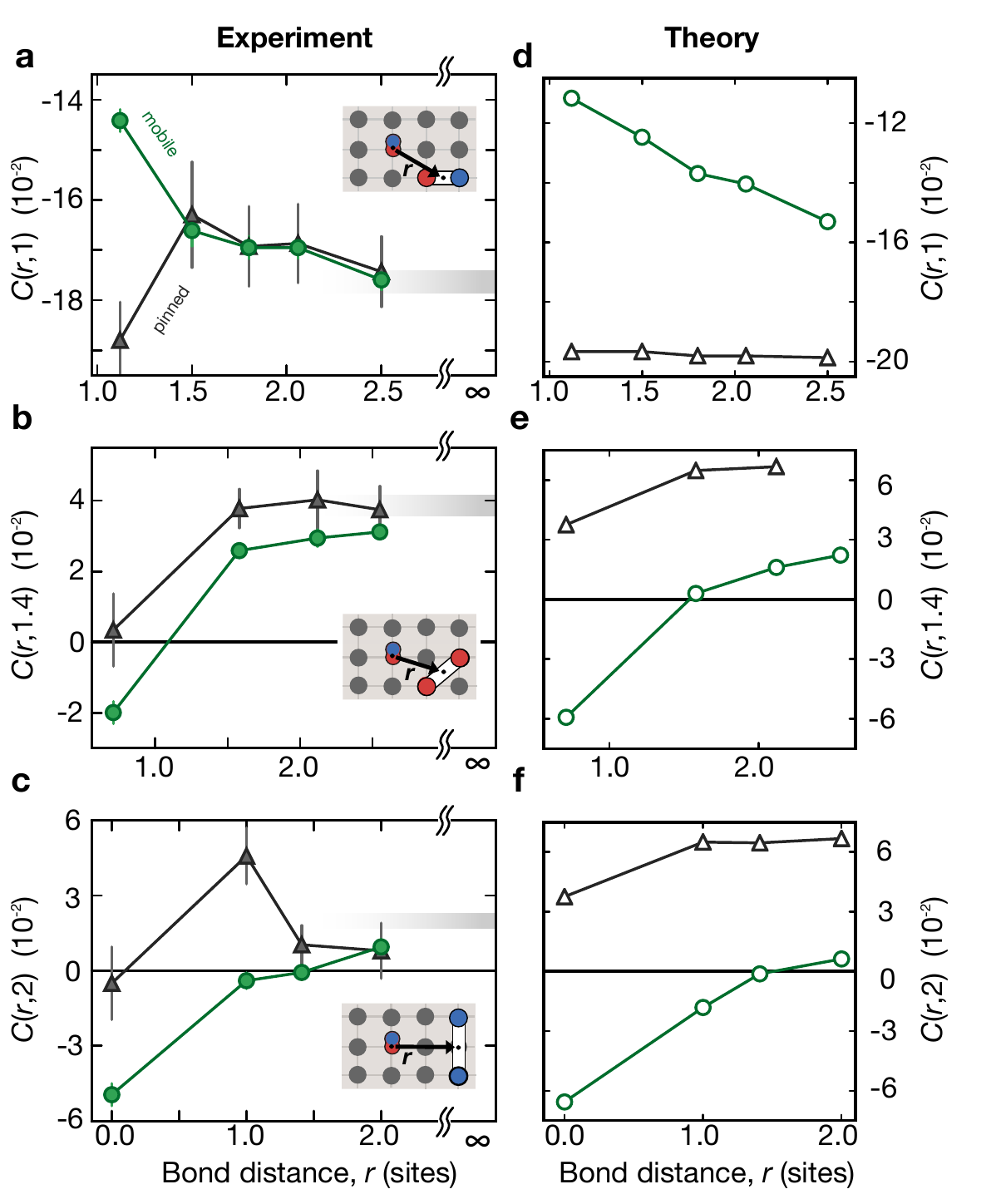}
\caption{\textbf{Spin correlations as a function of bond distance from doublons. }
Comparison between experiment (left panel) and numerical calculations of a string model for magnetic polarons and exact diagonalization of an immobile doublon in the $t$-$J$ model (right panel). \textbf{a}, \textbf{d}, Nearest-neighbour (NN), \textbf{b}, \textbf{e}, diagonal and \textbf{c}, \textbf{f}, next-nearest-neighbour (NNN) spin correlations as a function of bond distance (see inset) from mobile (green) or immobile (black) doublons. Error bars denote one s.e.m. For mobile doublons, diagonal and NNN correlations within an average bond distance of one lattice site are sign flipped w.r.t the AFM background. Correlations quickly recover at larger distances to a value approaching the undoped antiferromagnetic value, represented by a grey bar at a distance 'infinity' with a width of two s.e.m. The amplitude of the correlation changes with bond distance is well captured by the string model (green theory). Also the sign reversal of diagonal and NNN correlations is predicted. In stark contrast to the mobile case, for pinned doublons a slight enhancement of NN and NNN correlations around a distance of one site is visible in the experiment. Correlations between neighbouring spins of the trapped doublon are weakend, which is consistent with exact diagonalization calculations (black theory). \label{fig:fig4}
}
\end{figure}

\section*{Tweezer}
To demonstrate that the mobility of the doublon, i.e. doublon delocalization, is a key ingredient for polaron formation, we now investigate the effect of an artificially introduced localized doublon on the surrounding magnetic correlations. This situation is akin to doping nonmagnetic defects in solid state systems \cite{Alloul2009}. We set the chemical potential to prepare a system without doping and adiabatically ramp up the power of an optical tweezer focused to a single central site, while simultaneously ramping up the lattice. The final tweezer depth was set such that the density of that site saturates at 1.77(1) (see Fig. \ref{fig:fig3}a). A perfectly deterministic doublon preparation is not reached in our experiments, most likely due to detection errors and higher band effects (see supplement). We analyze the same doublon-conditioned three-point correlator $C(\textbf{r}_0;\textbf{r},\textbf{d})$ for diagonal spin correlations as before, with $\textbf{r}_0$ fixed to the pinned site (see Fig. \ref{fig:fig3}b). As expected, the strong spin distortion is absent in this case. Instead, magnetic correlations across the trapped site are slightly reduced in correlation compared to the undoped background (see Fig. \ref{fig:fig3}c). On the other hand, lattice sites in the immediate vicinity of the trapped doublon exhibit significantly stronger antiferromagnetic correlations than the background, as will be shown in the next paragraph. 

\section*{Radial analysis and theory}
In order to allow for a quantitative study and a comparison to theoretical models, we group the three-point spin correlations by the magnitude of their bond distance $r=|\textbf{r}|$ from doublons. Measured NN, diagonal and NNN spin correlations are shown in Fig. \ref{fig:fig4}. The local distortion of spin correlations around mobile doublons is visible in all correlators. Sign reversal of diagonal and NNN correlations occurs at a mean bond distance of one site, yielding a diameter and estimate for the polaron size of around two lattice sites. An enhancement of spin correlations around the pinned site is visible in the closest NN correlations and distance one NNN correlations. We compare our findings to theoretical model calculations, carried out for the estimated temperature of our system (see Fig. \ref{fig:fig4}). For the mobile case, an effective string model of magnetic polarons is used, based on the assumption of frozen-out spin dynamics. Remarkably, similar amplitude changes of correlations and hence a similar polaron radius is predicted (also found in exact diagonalization results of the $t$-$J$ model on a 4x4 system, see supplement). Furthermore, the sign changes of correlations in the vicinity of the doublon are reproduced in this model as seen also in Fig \ref{fig:fig4} e,f. Quantitative differences between the effective model and the experiment remain, which, however, can be expected due to the only moderate separation of spin and hole dynamics ($J/t=0.3$) and the elevated temperatures in the experimental system. In the case of pinned doublons, exact diagonalization calculation of the $t$-$J$ model is performed with zero tunneling of the excess doublon. For this calculation, the weakening of correlations between spins to the left and right (or below and above) of the doublon is reproduced. Since the pinned doublon blocks the path connecting these spins, these correlations are weakened. The enhancement effect observed in the experiment is not captured by our calculation at finite temperature and absolute correlation values from exact diagonalization suffer from strong finite size effects. We note that the observed effect of enhanced correlations is in fact opposite to the local spin distortion in the case of of mobile doublons.


\section*{Conclusion and Outlook}
We presented single-particle resolved imaging of a magnetic polaron in a doped Fermi-Hubbard system, by revealing the dressing of mobile doublons with a spin distortion. Artificially localizing the doublon made the polaronic spin distortion disappear, as the competition between kinetic and magnetic energy is suppressed. In the future, effective mass or quasiparticle weight of the polaron could be probed by transport \cite{Brown2018, Nichols2018} or spectroscopic methods \cite{Stewart2008}. By implementing larger and more homogeneous systems as well as new cooling schemes \cite{Lubasch2011, Kantian2016, Mazurenko2017}, a microscopic study of the crossover from polarons to the emergence of pseudogap, strange metal, stripe phases or pairing is within reach. Our full resolution of density and spin gave access to an entirely new characterization technique of polarons and provides new observables for the exploration of strongly correlated phenomena and their microscopic origin. 

\bigskip

\begin{acknowledgments}
\textbf{Acknowledgments:} We thank D. Stamper-Kurn, Yao Wang, Efstratios Manousakis, Subir Sachdev and Thierry Giamarchi for fruitful discussions. This work benefited from financial support by the Max Planck Society (MPG) and the European Union (UQUAM, FET-Flag 817482, PASQUANS). J.K. gratefully acknowledges funding from Hector Fellow Academy. F.G. and E.D. acknowledge support from Harvard-MIT CUA, NSF Grant No. DMR-1308435, AFOSR-MURI Quantum Phases of Matter (grant FA9550-14-1-0035). F.G. acknowledges financial support from the Gordon and Betty Moore foundation through the EPiQS program and from the Technical University of Munich - Institute for Advanced Study, funded by the German Excellence Initiative and the European Union FP7 under grant agreement 291763, from the DFG grant No. KN 1254/1-1, and DFG
TRR80 (Project F8).
\end{acknowledgments}

\bigskip

\textbf{Materials and Correspondence:} Correspondence and requests for
materials should be addressed to joannis.koepsell@mpq.mpg.de



%

\newpage

\section*{Supplementary Information:}
\setcounter{figure}{0}
\renewcommand\thefigure{S\arabic{figure}}  
\subsection{Experimental sequence}
The preparation of cold Fermi-Hubbard systems closely followed the procedure described in Salomon \textit{et al.}\cite{Salomon2018}. We started by preparing a balanced mixture of the two lowest hyperfine states of fermionic $^{6}$Li ($F=1/2\,;m_{F}=\pm1/2$), harmonically confined in a single two-dimensional plane of a $40\,E_{r}^ {z}$ deep optical lattice with $3.1\,\mu$m spacing in $z$-direction. The final atom number was set by the evaporation parameters. Subsequently, we ramped the $x$ and $y$-lattice with spacings $a_{x}=1.15\, \mu$m and $a_{y}=2.3\, \mu$m linearly within $210\,$ms from 0 $\,E_{r}^{i}$ to their final values of $8.6\,E_{r}^{x}$ and $3\,E_{r}^{y}$. The final lattice depths were optimized with undoped Mott insulators to give strong and isotropic spin correlations. From band width calculations we extract the tight-binding nearest-neighbour tunneling amplitudes $t_{x}/h=170\,$Hz and $t_{y}/h=180\,$Hz. The next-nearest neighbour tunneling amplitude along $y$-direction is below $0.1 \times t_{y}$. Using the broad Feshbach resonance of $^{6}\text{Li}$, the scattering length was tuned during the lattice ramp from $350\,a_{B}$ to $2150\,a_{B}$ in two linear ramps. The first one ramped to $980\,a_{B}$ within $150\,$ms and the second ramp increased the scattering length to $2150\,a_{B}$ in $60\,$ms. Applying a local Stern-Gerlach detection technique \cite{Boll2016} and subsequent Raman sideband cooling in a pinning lattice\cite{Omran2015}, the local spin and density on each lattice site was obtained with an average fidelity of $97\%$. 

\begin{figure}[b]
\centering
\includegraphics{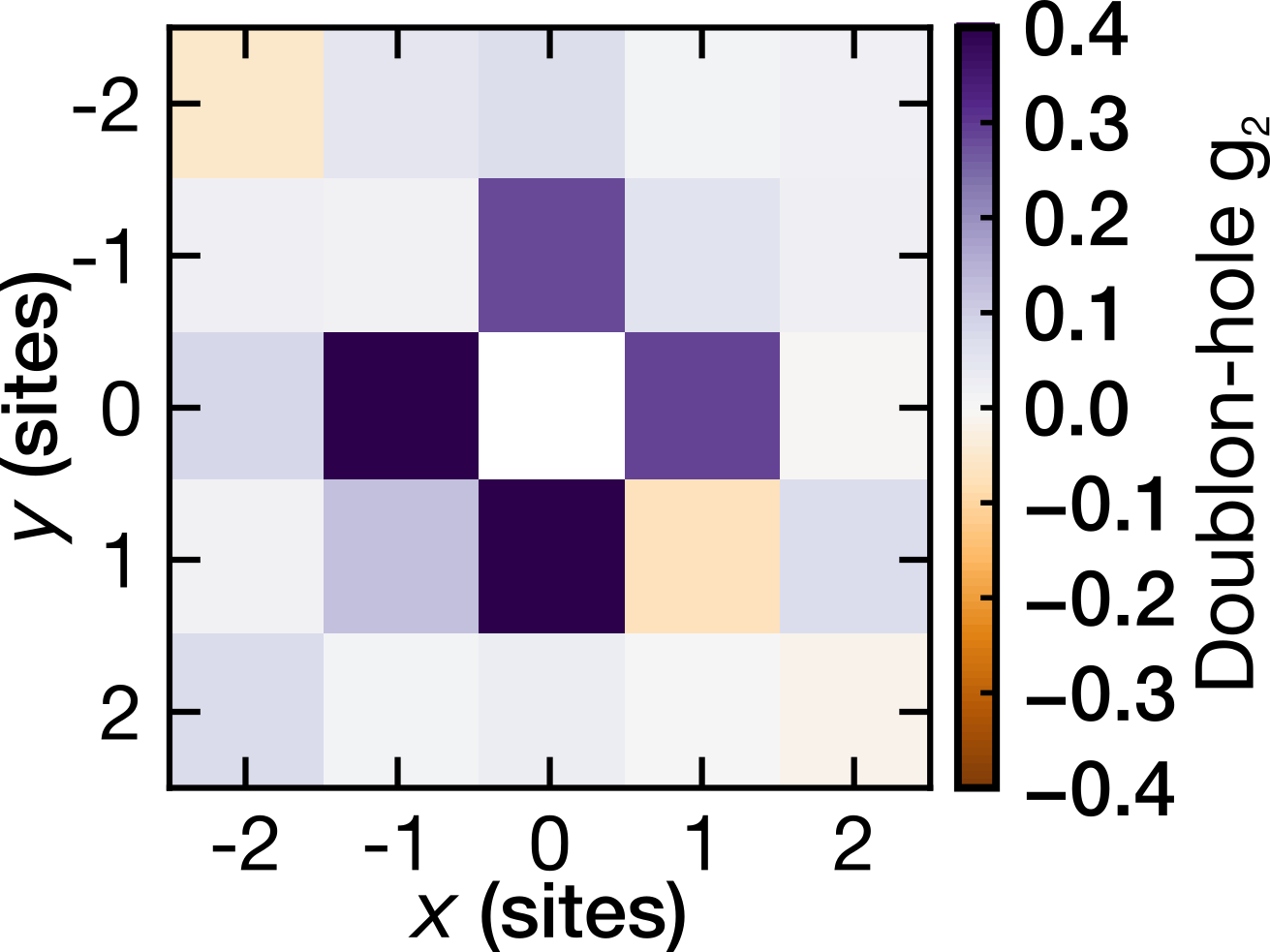}
\caption{
\textbf{Doublon-hole correlation.}
Two-point $g_2$ correlation function between double occupations and holes, showing a strong bunching effect at nearest-neighbour distances. This motivates our procedure to neglect double occupations with holes as nearest-neighbours for the analysis of mobile doublons.
\label{fig:s3}
}
\end{figure}

\subsection{Data analysis}
\begin{figure*}
\centering
\includegraphics{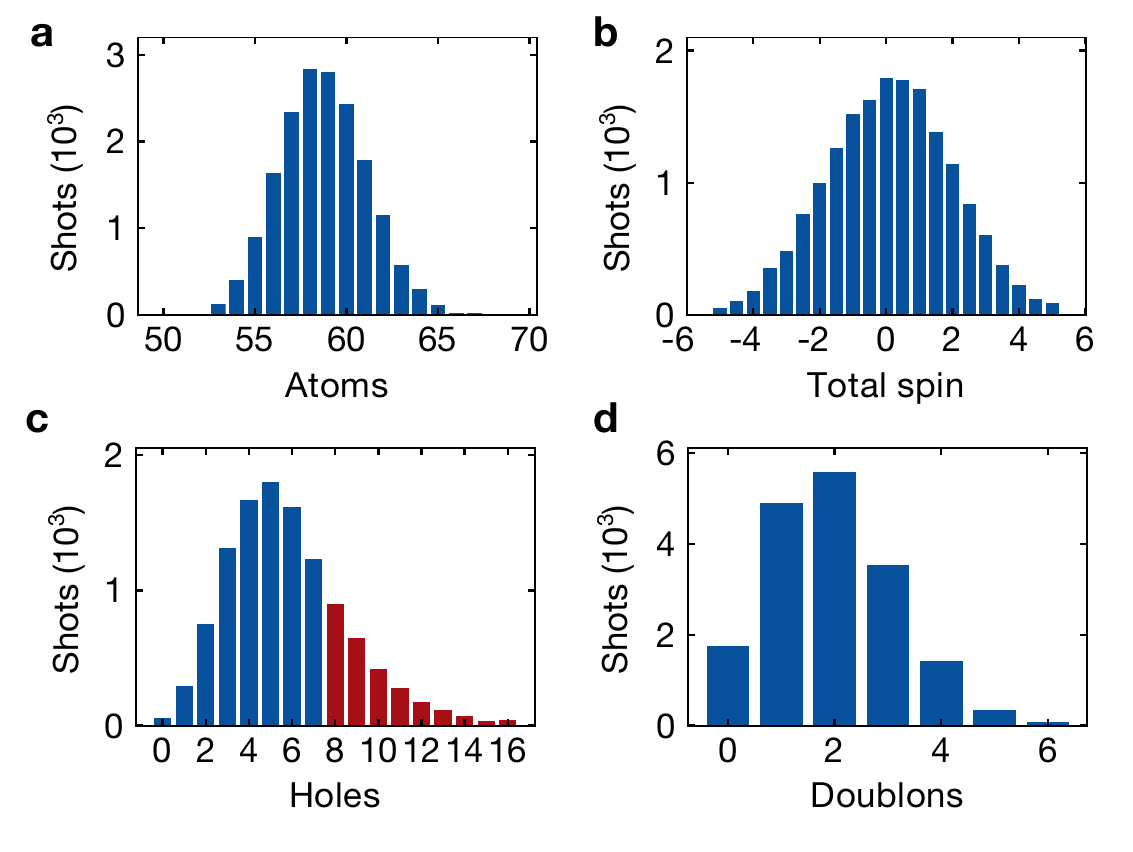}
\caption{
\textbf{Dataset statistics for measurement of mobile doublon.}
Histogram of the number of \textbf{a}, atoms \textbf{b}, spins and \textbf{c}, holes in the region with density greater than $0.7$. Red bars in \textbf{c} indicate discarded shots by the applied hole filter (see text). \textbf{d}, Histogram of the number of mobile doublons (doublon-hole fluctuations subtracted) in the doped $5 \times 3$ site region.
\label{fig:s4}
}
\end{figure*}

The work presented here is based on a dataset for pinned and one for mobile doublons. For the mobile and for the pinned case,  measurements with 33,669 and 9,002 images were taken. In the analysis, we considered only shots with total spin $|S^{z}_{\text{tot}}| \leq 3.5$, in order to filter out fluctuations in our spin detection scheme\cite{Salomon2018} and stronlgy magnetized clouds. This corresponds to a maximum allowed magnetization of $|S^{z}_{\text{tot}}|/N \simeq 0.05$ and approximately $68\%$ of the total images recorded. All further analysis was performed on lattice sites with a mean density $n\geq0.7$. This region corresponds to the sites shown in Fig. 2a and 3a both for the mobile and pinned cases respectively. In order to exclude clouds heated from inelastic three-body collisions, images with a total number of holes $\sum\limits_{ROI} n_h \geq 8$ contained in this ROI were discarded (amounting to neglecting around $16\%$ of the data). Computing the doublon-hole correlation function $g_{2}=\langle \hat{n}_d \hat{n}_h \rangle /(\langle \hat{n}_d \rangle \langle \hat{n}_h \rangle)-1$ reveals doublon-hole bunching on nearest-neighbour distances, see Fig. \ref{fig:s3}, which indicates the presence of doublon-hole fluctuations.  In order to distinguish between mobile doublons and doublon-hole fluctuations, we excluded doublons with holes as nearest-neighbours from the analysis. The resulting dataset statistics after processing is shown in Fig. \ref{fig:s4} for the mobile case.

\subsection{Chemical potential calibration}
\begin{figure}[t]
\centering
\includegraphics{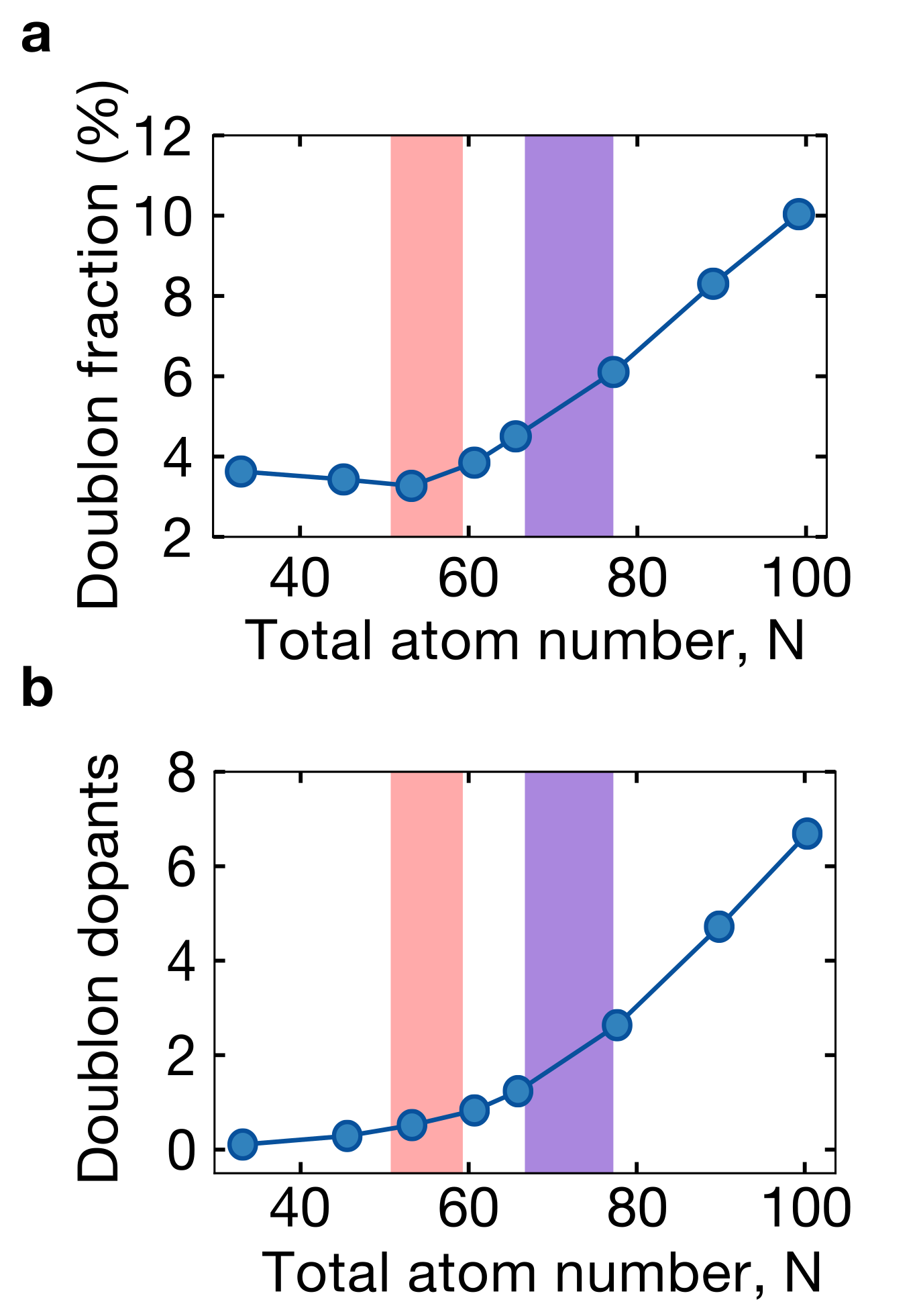}
\caption{
\textbf{Calibration of the chemical potential.}
\textbf{a}, Fraction of double occupations and \textbf{b}, number of doped doublons (excluding doublon-hole fluctuations) in the system as a function of total atom number $N$. Statistical error bars are smaller than the point size. Pinned doublon measurements were taken in an undoped system (red bar). For the mobile doublon dataset, settings for weak doping were used (purple bar). The bar width represents the standard deviation obtained from atom number fluctuations.
\label{fig:s1}
}
\end{figure}

In order to control the doping, we measured the number of double occupations per number of atoms ($\frac{N_{\text{doub}}}{N}$) in our system as a function of the atom number $N$ (see Fig. \ref{fig:s1}a). For low atom numbers, the doublon fraction saturates below $4\%$, which we attribute to quantum fluctuations in the form of doublon-hole pairs. The background of doublon-hole fluctuations is confirmed by discarding doublons with holes as nearest-neighbours, obtaining the curve of doped doublons versus total atom number in Fig. \ref{fig:s1}b. For low atom numbers, no doped doublons are present, while at higher atom numbers finite doping sets in. To probe individual mobile doublons in the Fermi-Hubbard model, we used systems with about 72 atoms and around $2$ doped doublons. To study the effect of localized doublons, created with an optical tweezer (see below), we used smaller systems with around 55 atoms to avoid the effect of doping.

\begin{figure}
\centering
\includegraphics{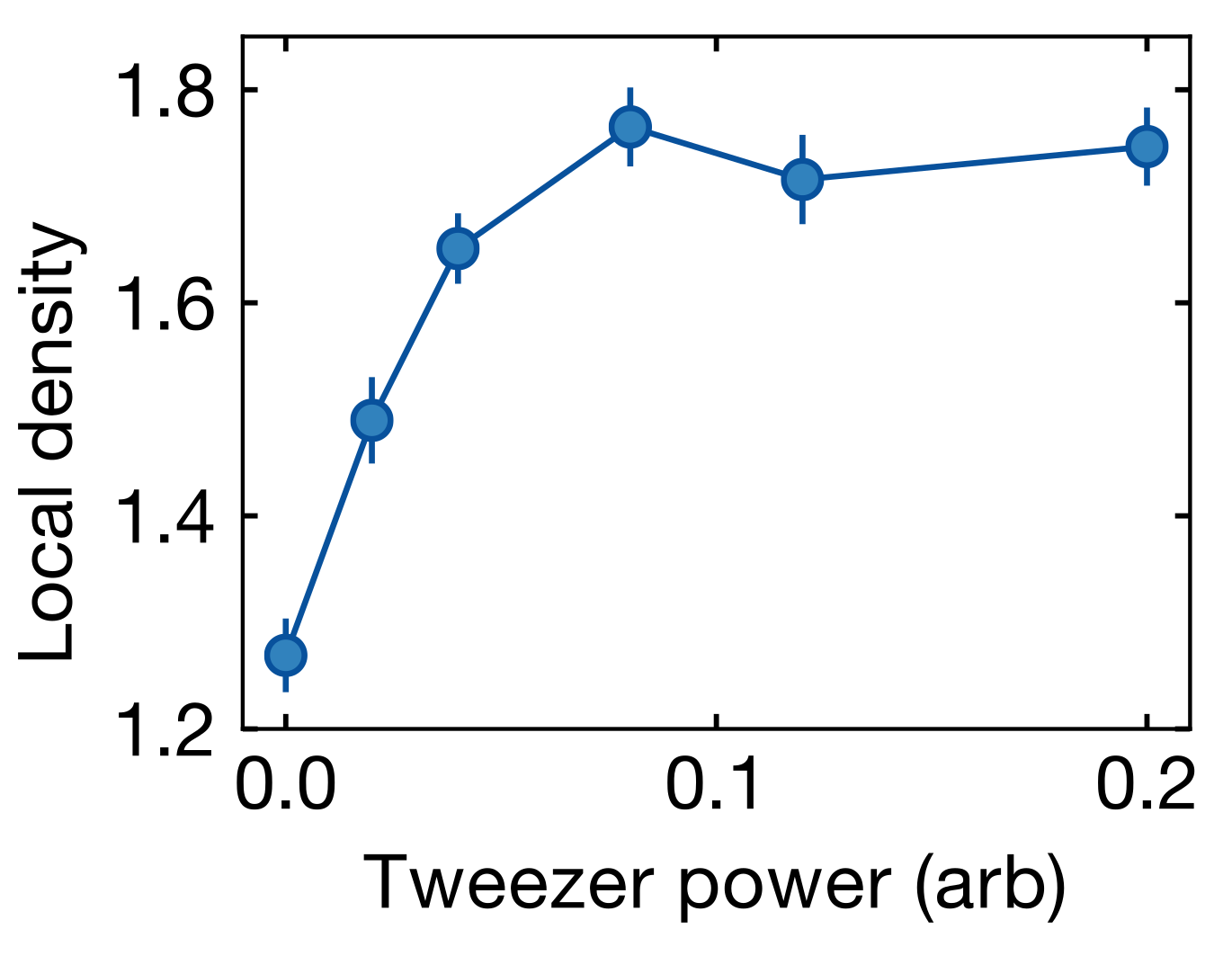}
\caption{
\textbf{Calibration of tweezer power.}
Density of the lattice site, on which the tweezer is focused, as a function of final tweezer power. Error bars denote one s.e.m. For the realization of pinned doublons the power was set to 0.11 (arbitrary units). 
\label{fig:s2}
}
\end{figure}

\subsection{Tweezer depth calibration}
For the pinned doublon case, we ramped the power of an optical tweezer beam focused on a single lattice site to its final value simultaneously with the $xy$-lattice depth. The tweezer depth was calibrated in a separate measurement by determining the density of the target site as a function of the final tweezer power. As shown in Fig. \ref{fig:s2}, the density first increases with power and then saturates below $1.8$, independent of higher final powers. The total detected local density $n$ of $1.77$ is composed of $n=3 \cdot n_t + 2\cdot n_d + 1\cdot n_s +0\cdot n_h$, where $n_t$, $n_d$, $n_s$ and $n_h$ are the triplon-, doublon-, singlon- and hole-density. For our measurement of localized doublons, we set the tweezer depth to the value, where the density starts saturating. At this tweezer power, the hole density $n_h$ at that site is $0.07$, the singlon density $n_s$ is $0.13$, the doublon density $n_d$ is $0.74$ and a small triplon density $n_t$ of $0.05$ exists, which we attribute to imperfections in the detection of doublons, imperfections in the loading procedure and coupling to higher bands of the $y$-direction. In combination with our finite imaging fidelity of $97\%$, this explains why a deterministic preparation of the doublon is not fully achieved.

\begin{figure}
\centering
\includegraphics{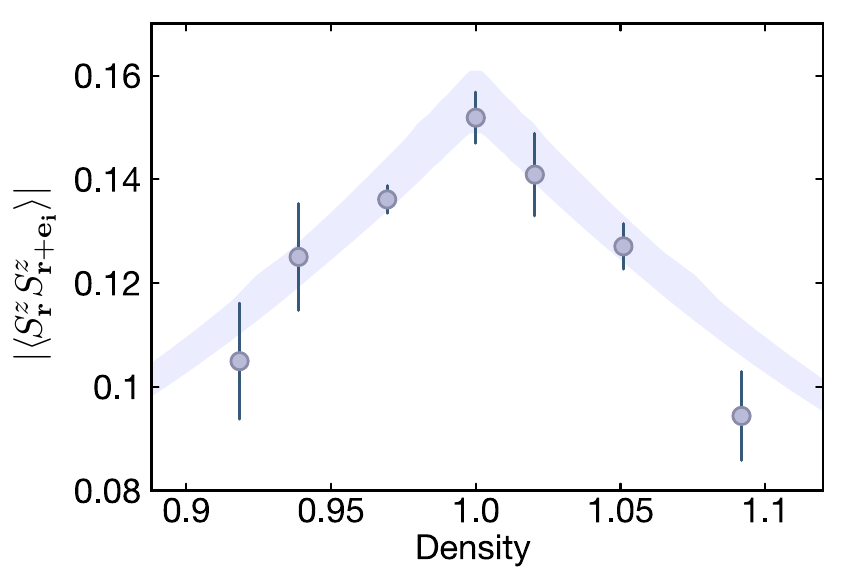}
\caption{
\textbf{Temperature estimation.}
Two-point nearest-neighbour spin correlations as a function of binned density. Error bars denote one s.e.m. Upper (lower) values of the indicated blue band correspond to temperatures of $T/t=0.43$ ($0.46$) in  numerical linked-cluster expansions at $U/t=13$.
\label{fig:s7}
}
\end{figure}

\begin{figure*}
\centering
\includegraphics{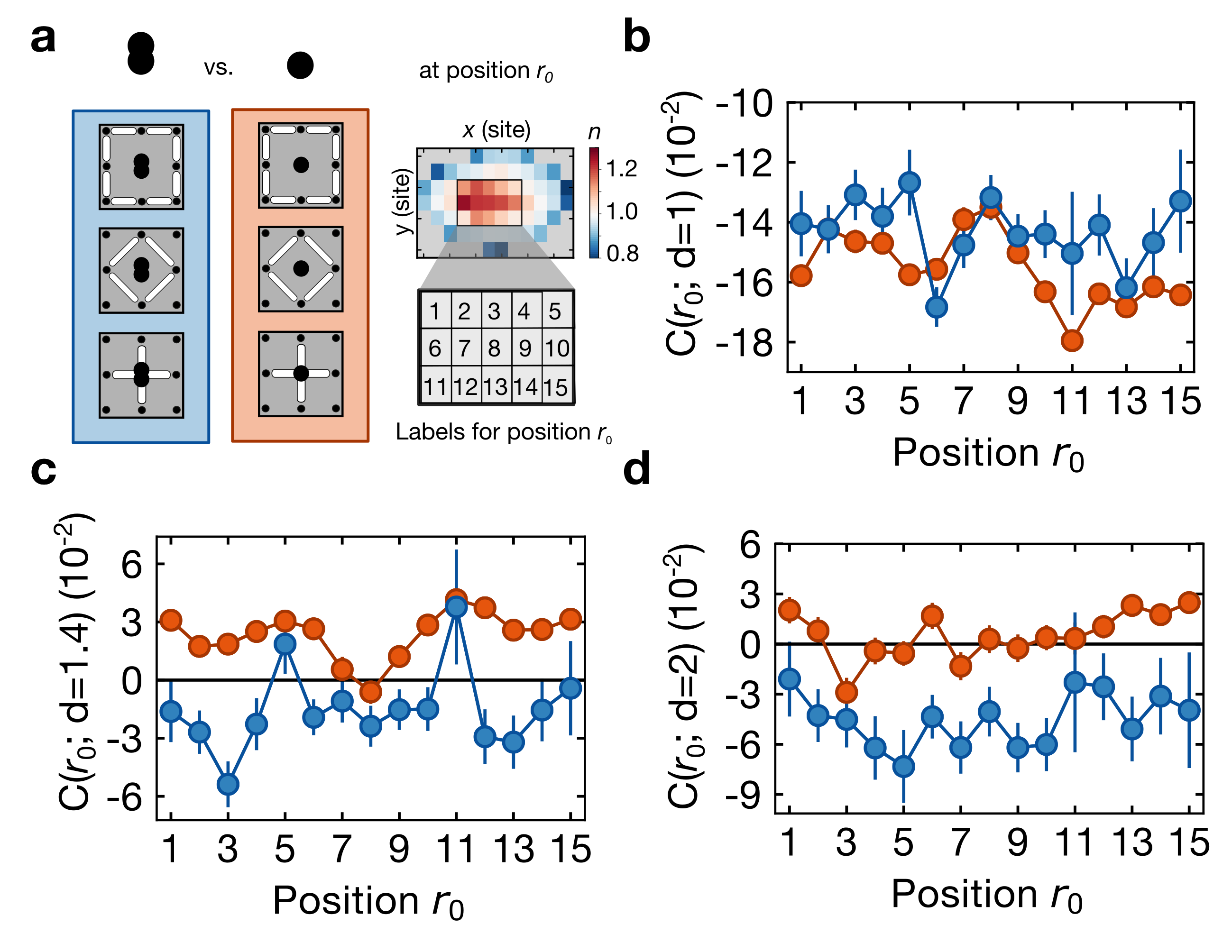}
\caption{
\textbf{Extended polaron analysis.}
\textbf{a}, Comparison of the local spin environment around a lattice site $\textbf{r}_0$, if it is occupied by a doublon (double black circle) or singlon (single black circle). To simplify notation, site positions in the doped region of our system are labeled from $0$ to $15$ (see inset in density distribution). The spin distortion is present in \textbf{b}, nearest-neighbour \textbf{c}, diagonal and \textbf{d}, next-nearest-neighbour correlations whenever a doublon is detected (blue) and absent whenever a singlon is detected (red) at any position in the $5 \times 3$ site system. Error bars denote one s.e.m. Next-nearest-neighbour correlations are measured across doublons and have a strong signal-to-noise ratio, which we attribute to their short bond distance of $0$, compared to e.g. nearest-neighbour correlations (bond distance $1.1$). 
\label{fig:s5}
}
\end{figure*}

\subsection{Temperature estimation}
To estimate the temperature of the clouds, we compared the loss-corrected nearest-neighbour spin correlations $\langle S^z_\mathbf{r}S^z_{\mathbf{r+e_{i}}}\rangle$, where $\mathbf{e_{i}}=e_x,\,e_y$, close to half filling with numerical linked-cluster expansions (NLCE) up to ninth order for homogeneous systems \cite{Khatami2011}. We used Wynn's algorithm \cite{Rigol2007} to sum the terms of the series and obtain nearest-neighbor spin correlations as a function of the density for $U/t=13$. This value, slightly lower than the one stated in the main text, takes into account the renormalization of the interaction strength $U$ for low lattice depths \cite{Buchler2010}.
The experimental spin correlations as a function of density were obtained by averaging over sites with local densities between 0.9 and 1.1 in bins ranging from 0.02 to 0.04 to collect enough statistics.
We find that our experimental correlations compare well with NLCE results at a temperature $T/t=0.45(4)$ (see Fig. \ref{fig:s7}).

\begin{figure*}
\centering
\includegraphics{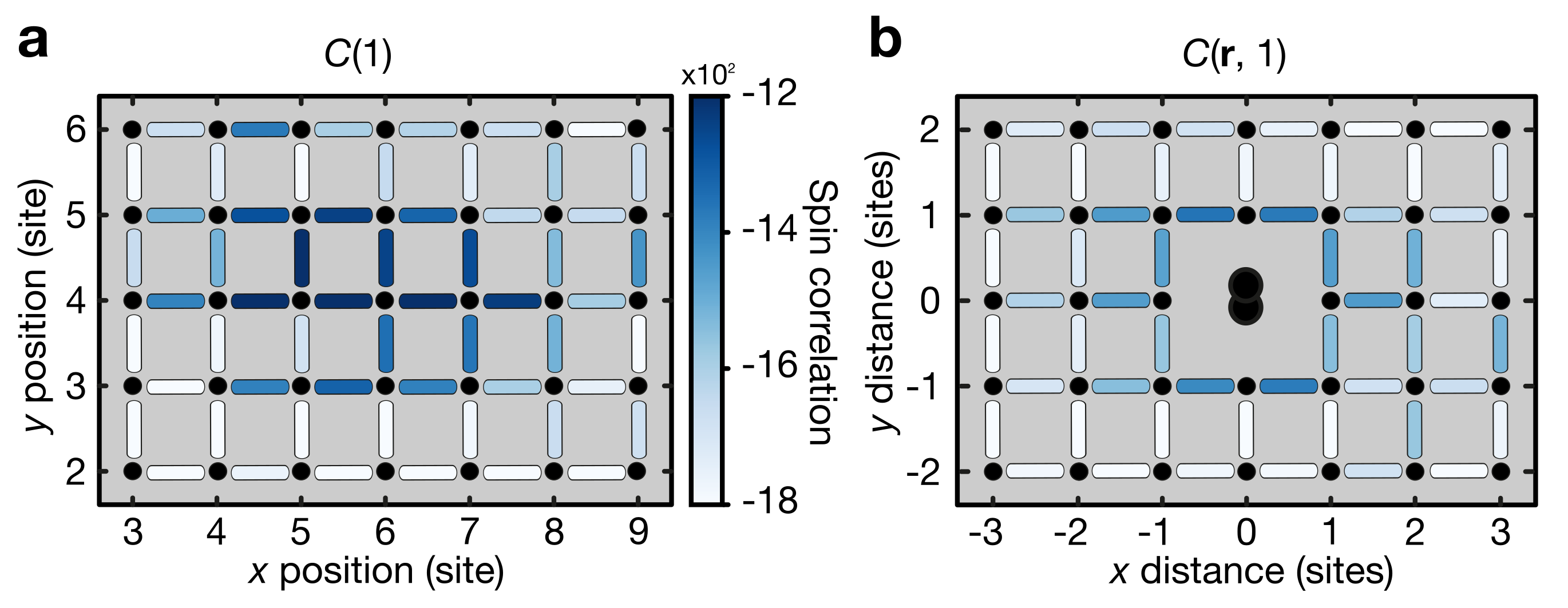}
\caption{
\textbf{Nearest-neighbour correlations with mobile doublons.}
\textbf{a}, Two-point nearest-neighbour correlations in the central region, represented by colored bonds between lattice sites (black dots). \textbf{b}, Nearest-neighbour spin correlations as a function of distance from mobile doublons.
\label{fig:s6}
}
\end{figure*}

\subsection{Extended polaron analysis}
The spin correlator $C(\textbf{r},\textbf{d})$ discussed in the main text was used to reveal the polaronic spin environment of mobile doublons. This correlator is the result of averaging $C(\textbf{r}_0;\textbf{r},\textbf{d})$  over positions $\textbf{r}_0$ of mobile doublons. Here we show, that the spin distortion is consistently dressing the doublon independent of detected position in the trap. We study the nearest-neighbour (NN), diagonal and next-nearest-neighbour (NNN) correlations with the shortest bond distance $\textbf{r}$ to the doublon as a function of position $\textbf{r}_0$. In order to maintain a sufficiently high signal-to-noise ratio, we average bonds isotropically to obtain a single value. Furthermore, we contrast this to the case , when position $\textbf{r}_0$ is singly occupied instead
\begin{align}
 C_{\text{singlon}}(\textbf{r}_0;\textbf{r},\textbf{d} ) & \equiv C_{\text{singlon}}(\textbf{r}_0;\textbf{r}_1,\textbf{r}_2 ) \nonumber \\
 &=4 \langle S^z_{\textbf{r}_1} S^z_{\textbf{r}_2}  \rangle_{ \scalebox{0.65}{\newmoon}_{\textbf{r}_0}  \scalebox{0.65}{\newmoon}_{\textbf{r}_1}\scalebox{0.65}{\newmoon}_{\textbf{r}_2}}.
\end{align}
The total correlation strength for those two different cases is shown in Fig. \ref{fig:s5} as a function of position $\textbf{r}_0$ in the system for NN, diagonal and NNN correlations. A strong difference in the local spin environment is observed, depending on if a doublon or singlon is present at a specific site. The spin distortion that dresses the doublon is strongest for NNN correlations and weakest for NN correlations, which can be understood by the fact that NNN correlations are much closer (bond distance $=0$) to the doublon than NN correlations (bond distance $=1.1$). When a singlon occupies a certain position, the strong spin distortion is absent. In this case the spin correlation surrounding it does not fully return to the background value of an undoped system, because polarons are still present in the system and their average distance from the singlon is on the order of one to two lattice sites. This fact is also responsible for the varying correlation strength for singlons at different positions. When the singlon is considered in regions of higher density, the average distance to polarons decreases, leading to a parasitic reduction in correlation strength. 

\subsection{Nearest-neighbour correlations}
\begin{figure}
\centering
\includegraphics{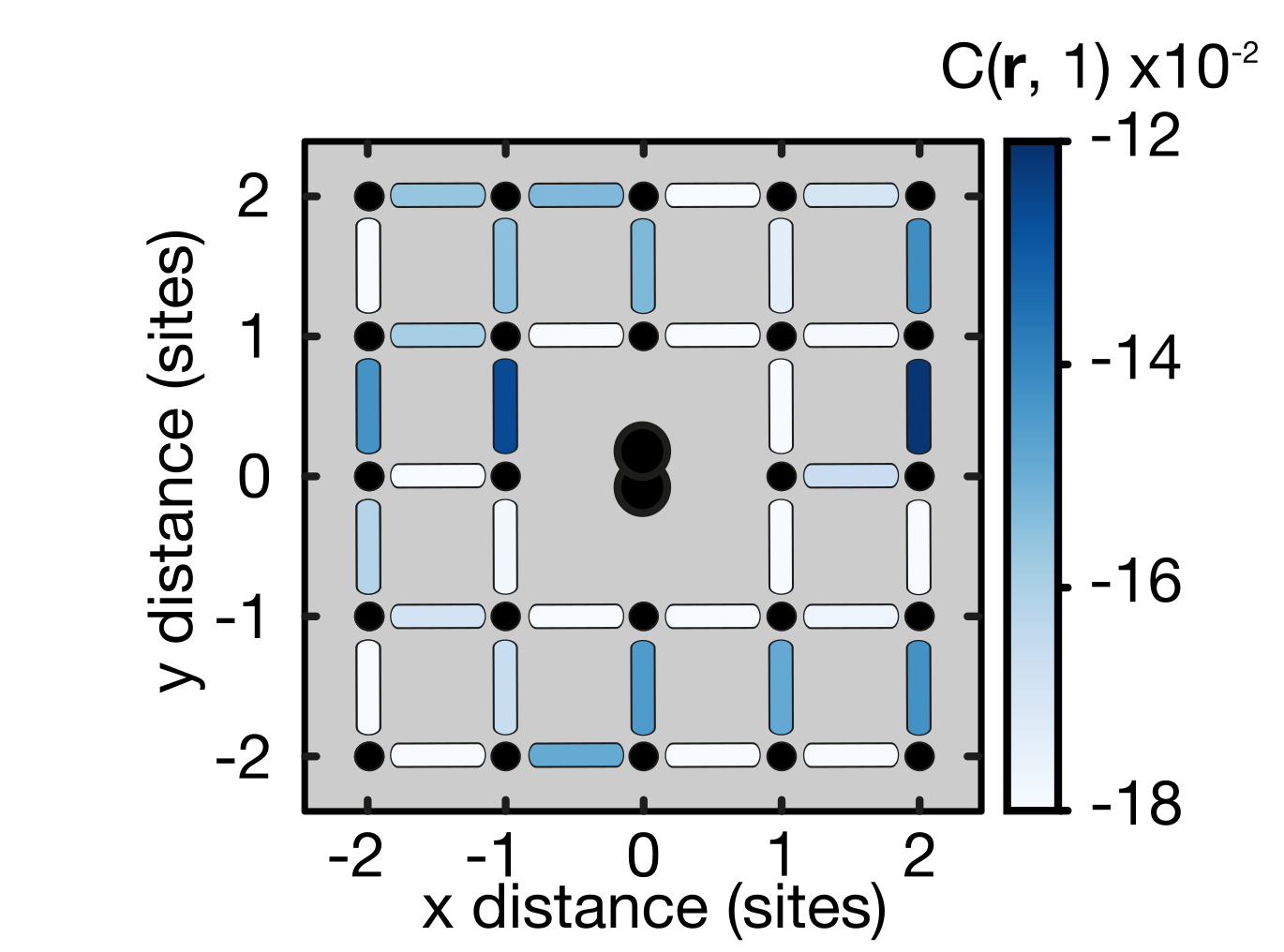}
\caption{
\textbf{Nearest-neighbour correlations around pinned doublons.}
The enhancement effect of correlations next to pinned doublons is visible in the strong bonds surrounding the trapped site.
\label{fig:s8}
}
\end{figure}

In analogy to diagonal spin correlations in the main text, NN correlations (c.f. eq. 1 and 2 in the main text with $|\textbf{d}|=1$) are shown in Fig. \ref{fig:s6} globally and around mobile doublons. The spin distortion dressing the doublon is also visible here. Nonetheless, as explained above, the signal to noise ratio is weaker than for the other correlators. In Fig. \ref{fig:s8} NN correlations are shown in the case of pinned doublons. The local enhancement of correlations is visible at the closest bond distance.

\subsection{Diagonalization calculations}
In the main text we compare our experimental results to numerical simulations of a single doublon. Starting from the Fermi-Hubbard model with a single doublon and  performing an expansion to leading order in $t/U$, the so-called $t-J^*$ model is obtained \cite{Auerbach1998}
\begin{multline}
\hat{\mathcal{H}}= - \frac{J}{2} \sum_{\langle \textbf{i}, \textbf{j} \rangle} \left[ \l \hat{\mathcal{A}}^\dagger_{{\langle \textbf{i}, \textbf{j} \rangle}} \hat{\mathcal{A}}_{{\langle \textbf{i}, \textbf{j} \rangle}} - 1/2  \r  \l 1 - \hat{d}^\dagger_{\textbf{j}} \hat{d}_{\textbf{j}} - \hat{d}^\dagger_{\textbf{i}} \hat{d}_{\textbf{i}} \r \right] +\\
+ t \sum_{\langle \textbf{i}, \textbf{j} \rangle} \left[  \hat{d}^\dagger_{\textbf{i}} \hat{d}_{\textbf{j}} \hat{\mathcal{F}}^\dagger_{{\langle \textbf{i}, \textbf{j} \rangle}}  + \text{h.c.} \right] + \frac{J}{4} \sum_{\langle \textbf{i}, \textbf{j}, \textbf{k} \rangle}  \hat{d}^\dagger_{\textbf{k}} \hat{d}_{\textbf{i}} \hat{\mathcal{A}}^\dagger_{{\langle \textbf{i}, \textbf{j} \rangle}} \hat{\mathcal{A}}_{\langle \textbf{k}, \textbf{j} \rangle}.
\label{eqHtJoneHole}
\end{multline}
Here we expressed the localized magnetic moments $\hat{\textbf{S}}_{\textbf{j}}$ by Schwinger bosons $\hat{b}_{\textbf{j},\beta}$, such that $\hat{\textbf{S}}_{\textbf{j}} = \sum_{\alpha,\beta = \uparrow, \downarrow} \hat{b}^{\dagger}_{\textbf{j},\alpha} \boldsymbol{\sigma}_{\alpha,\beta} \hat{b}_{\textbf{j},\beta}$ , where $\boldsymbol{\sigma}_{\alpha,\beta}$ are matrix elements of Pauli matrices. The operators $\hat{d}_{\textbf{j}}$ describe the single doublon, using second quantization, and respect the constraint
\begin{equation}
\sum_{\alpha = \uparrow, \downarrow} \hat{b}^\dagger_{\textbf{j} \alpha} \hat{b}_{\textbf{j} \alpha}  + \hat{d}^\dagger_{\textbf{j}} \hat{d}_{\textbf{j}}= 1
\label{eq:constraintTJ}
\end{equation}
for all sites $\textbf{j}$ of the square lattice. The operators
\begin{flalign}
\hat{\mathcal{F}}_{{\langle \textbf{i}, \textbf{j} \rangle}} &= \hat{b}^\dagger_{\textbf{i} \uparrow} \hat{b}_{\textbf{j} \uparrow} +  \hat{b}^\dagger_{\textbf{i} \downarrow} \hat{b}_{\textbf{j} \downarrow},\\
\hat{\mathcal{A}}_{{\langle \textbf{i}, \textbf{j} \rangle}} &=  \hat{b}_{\textbf{i} \uparrow}  \hat{b}_{\textbf{j} \downarrow} -  \hat{b}_{\textbf{i} \downarrow}  \hat{b}_{\textbf{j} \uparrow},
\end{flalign}
are defined on a pair of neighboring sites ${\langle \textbf{i}, \textbf{j} \rangle}$, and $\langle \textbf{i}, \textbf{j}, \textbf{k} \rangle$ in Eq.~\eqref{eqHtJoneHole} denotes a sequence of three inequivalent neighboring sites (i.e. the case $\textbf{i}=\textbf{k}$ is not included).

The first term on the right-hand side of Eq.~\eqref{eqHtJoneHole} denotes spin-exchange interactions with magnitude $J = 4 t^2 / U$. The second term describes nearest-neighbor tunneling processes of the doublon, which lead to a distortion of the surrounding spins. The last term extends this $t-J$ model by three-site terms describing next-nearest neighbor hopping of the doublon correlated with exchange couplings of the two involved spins. A pinned doublon can be modeled by considering only the first term with amplitude $J \neq 0$ while setting $t=0$ and ignoring the last term in Eq.~\eqref{eqHtJoneHole}.

We implemented Eq.~\eqref{eqHtJoneHole} numerically on a $4 \times 4$ torus with periodic boundary conditions. We made use of the following conserved quantities, leading to a block-diagonal Hamiltonian: the $z$-component of the total conserved spin, $\hat{S}^z$ and the total conserved lattice momentum $\textbf{k}$ with quantized components $k_\mu = n_\mu \pi / 2$ for $n_\mu=0,1,2,3$ and $\mu=x,y$. To make the conservation of lattice momentum explicit, we use a discrete version of the Lee-Low-Pines unitary transformation~\cite{Lee1953}, which yields a spin Hamiltonian formulated in terms of Schwinger bosons where the doublon is localized in the origin. 

To calculate the properties of a thermal state at temperature $T = 1 / k_{\rm B} \beta$, we calculated the thermal density matrices $e^{- \beta \hat{\mathcal{H}}(S^z,\textbf{k})}$ separately in every sector and averaged over all sectors with the corresponding thermal weights $1/Z(S^z,\textbf{k})$, where $Z(S^z,\textbf{k}) = \text{tr} ~ e^{- \beta \hat{\mathcal{H}}(S^z,\textbf{k})}$. These calculations can be performed numerically exactly because the dimensionality of the largest Hilbert space with $S^z = \pm 1/2$ and arbitrary $\textbf{k}$ is only $6435$. 

In the main text we compare spin correlations between occupied sites, $\langle \hat{S}^z_{\textbf{i}} \hat{S}^z_{\textbf{j}} \rangle_{\scalebox{0.65}{\newmoon}_{\textbf{i}}\scalebox{0.65}{\newmoon}_{\textbf{j}}}$, directly to exact numerical results $\langle \hat{S}^z_{\textbf{i}} \hat{S}^z_{\textbf{j}} \rangle_{t-J}$ obtained within the $t-J^*$ model. This is justified because the restriction to singly occupied sites in the experimental data eliminates virtual doublon-hole pairs constituting the leading correction to the $t-J^*$ Hamiltonian.

\begin{figure}
\centering
\includegraphics{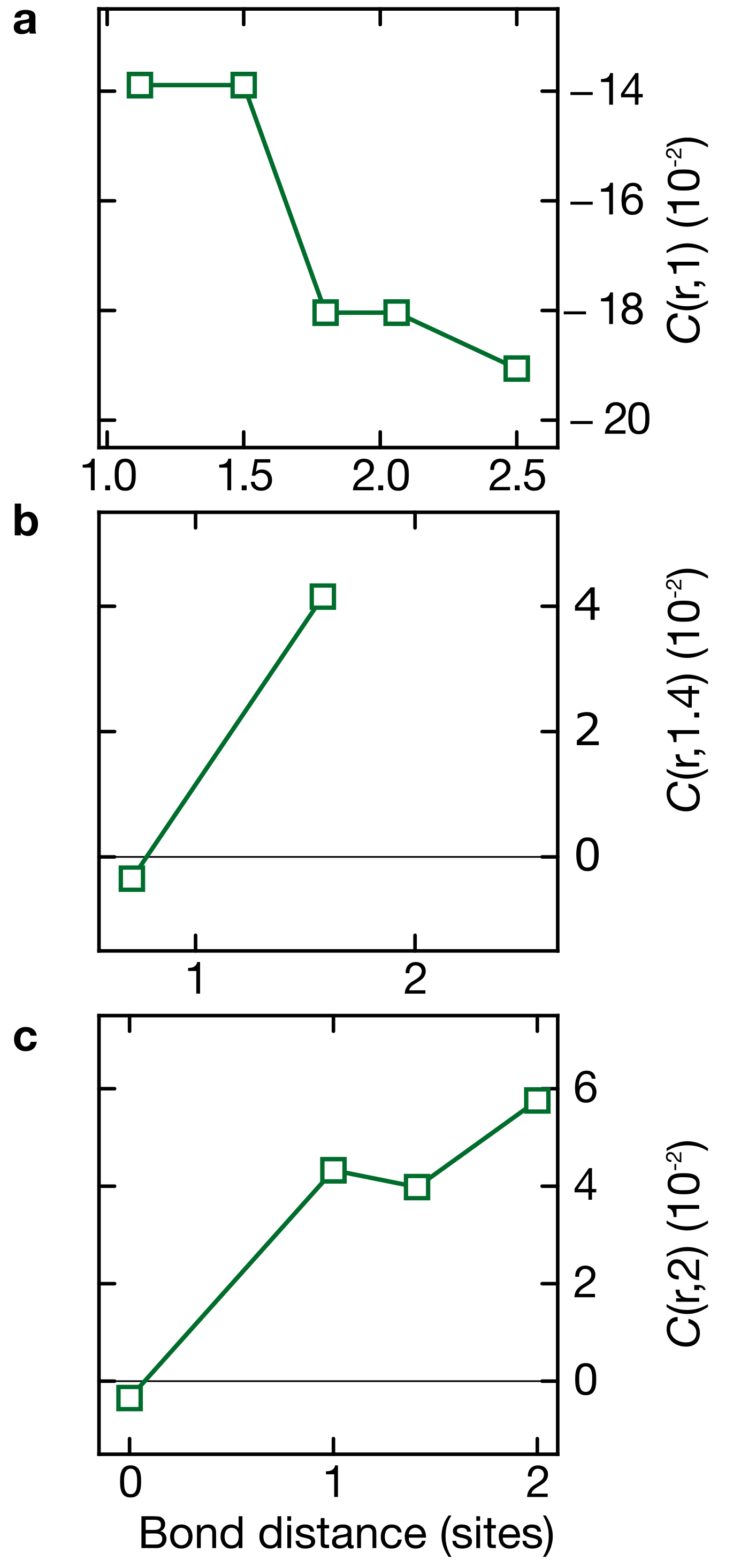}
\caption{
\textbf{Exact diagonalization results for mobile doublons.}
Spin correlations as a function of bond distance from mobile doublons for \textbf{a}, nearest-neighbour, \textbf{b}, diagonal and \textbf{c}, next-nearest-neighbour bonds.
\label{fig:s10}
}
\end{figure}

In Fig. \ref{fig:s10} we show our numerical exact diagonalization results of a mobile doublon. The amplitude and slope of correlation changes as a function of bond distance is very similar to the experimentally observed values (see main text). Also here, sign changes of diagonal and NNN correlations close to the doublon are visible. Absolute values differ from the experiment automatically by finite size offsets from the simulated $4 \times 4$ system.
~\\

\begin{figure}
\centering
\includegraphics{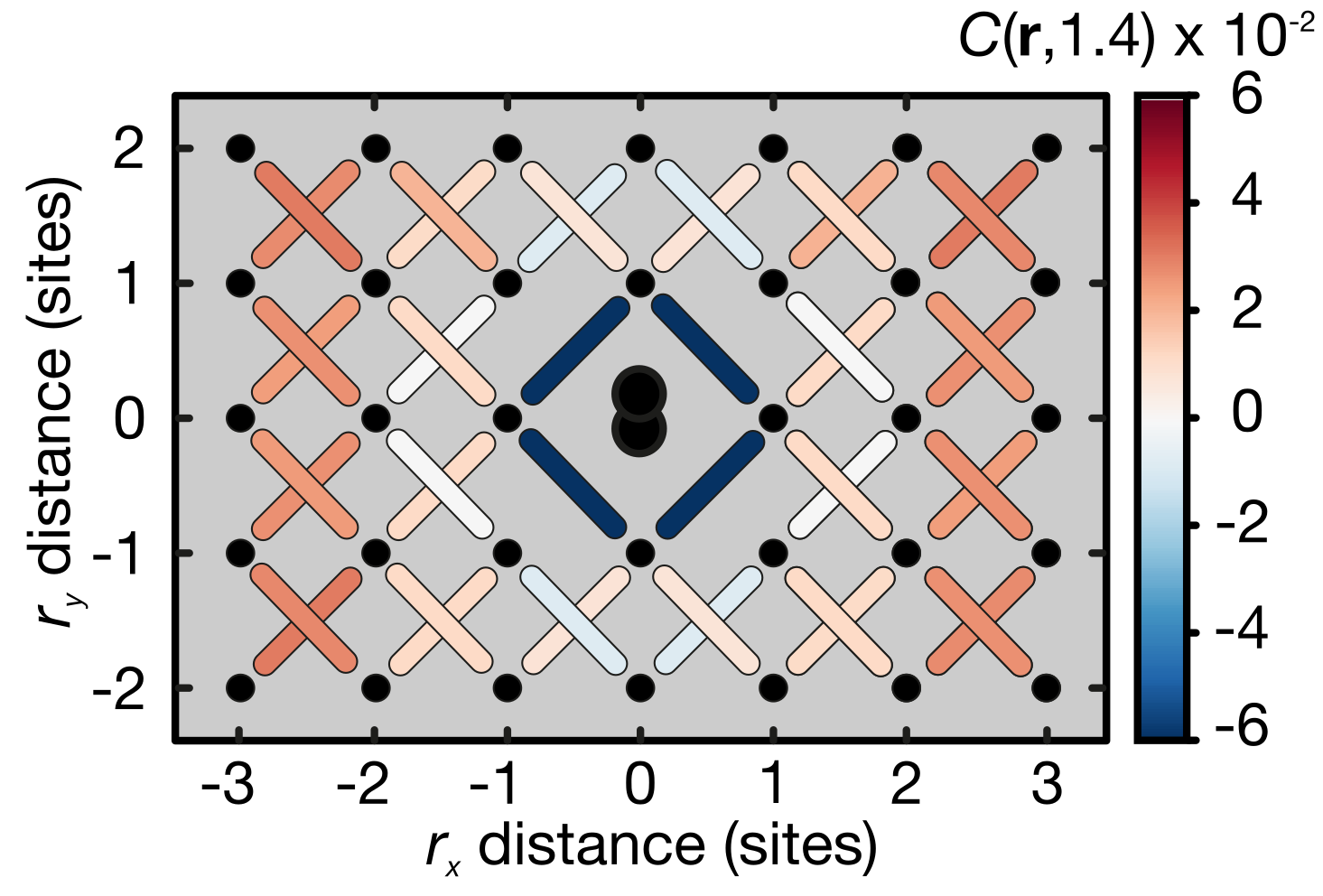}
\caption{
\textbf{Diagonal correlations around mobile doublons in the string model.}
Diagonal correlations (represented by colored bonds) between lattice sites (black dots) as a function of distance from mobile doublons (indicated in the center).
\label{fig:s9}
}
\end{figure}

\noindent
\subsection{Effective string model}
To obtain more insights into the qualitative signatures of magnetic polarons, and to include effects by the harmonic trapping potential, we discuss a simplified string model. Following Refs.~\cite{Grusdt2018,Grusdt2018mixD} we describe magnetic polarons as bound states of two partons: a spinon and a chargon. Such parton models of magnetic polarons were also discussed in different contexts in Refs.~\cite{Beran1996,Baskaran2007,Punk2015}. Diagonal spin correlations around mobile doublons obtained in this model are represented in Fig. \ref{fig:s9} in the same way as experimental correlations in the main text.

\emph{Hilbert space and Hamiltonian.--}
The effective spinon-chargon theory makes use of an approximate set of basis states $|{\textbf{j}^s,\Sigma,\alpha}\rangle$ labeled by the position of the spinon, $\textbf{j}^s$ in the square lattice and its spin $\alpha$, as well as the string configuration $\Sigma$ connecting the chargon to the spinon. The effective Hilbert space $\mathscr{H}$ factorizes,
\begin{equation}
\mathscr{H} = \mathscr{H}_s \otimes \mathscr{H}_\Sigma,
\end{equation}
with two parts: $\mathscr{H}_s$ describes the spinon and $ \mathscr{H}_\Sigma $ describes the string and, hence, the chargon. The string is always defined in the reference frame of the spinon.

Likewise we assume that the effective Hamiltonian contains two terms acting on the separate parts of the Hilbert space, as well as an interaction term,
\begin{equation}
\hat{\mathcal{H}}_{\rm eff} = \hat{\mathcal{H}}_s + \hat{\mathcal{H}}_\Sigma + \hat{\mathcal{H}}_{\rm int}.
\label{eqDefHspinonChargon}
\end{equation}
The first term describes the dispersion of the spinon $\omega_s(\textbf{k})$, which becomes
\begin{equation}
\hat{\mathcal{H}}_s = \sum_{\textbf{k},\alpha} \hat{s}^\dagger_{\textbf{k},\alpha} \hat{s}_{\textbf{k},\alpha} \omega_s(\textbf{k})
\end{equation}
in second quantization in momentum space. The second term describes fluctuations of the string configuration, reflecting the microscopic motion of the doublon in the $t-J$ model,
\begin{equation}
\hat{\mathcal{H}}_\Sigma = -  t \sum_{\langle \Sigma',\Sigma \rangle} |{\Sigma'}\rangle\langle{\Sigma}| + \text{h.c.}.
\end{equation}
Here a sum is taken over adjacent string configurations $\langle \Sigma',\Sigma \rangle$ which can be reached by a nearest-neighbor tunneling processes of the doublon. The interaction term describes a potential $V_{\Sigma}(\textbf{j}^s)$ for the string which depends on the spinon position,
\begin{equation}
\hat{\mathcal{H}}_{\rm int} = \sum_\Sigma \sum_{\textbf{j}^s, \alpha} \hat{s}^\dagger_{\textbf{j}^s,\alpha} \hat{s}_{\textbf{j}^s,\alpha}  ~ V_{\Sigma}(\textbf{j}^s) ~ |{\Sigma}\rangle \langle{\Sigma}|.
\end{equation}

The string potential contains two terms, $V_{\Sigma}(\textbf{j}^s) =  V_{1}(\textbf{j}_d) + V_{2}(\ell_\Sigma) $. The first term
\begin{equation}
V_{1}(\textbf{j}_d) = \frac{1}{2} \l W_x ( j_d^x )^2 + W_y ( j_d^y )^2 \r 
\end{equation}
contains the harmonic trapping potential seen by the doublon at position $\textbf{j}_d$, which is determined by combining $\textbf{j}^s$ and $\Sigma$; The strength of the trap in the experiment is around $W_x = 0.7 \,J$ and $W_y=2.6 \,J$. The second term describes the energy cost for distorting the spin environment when the chargon moves away from the spinon. This term is approximately linear in the string length $\ell_\Sigma$ and microscopic considerations show that it is proportional to the spin correlations $c_{\textbf{d}} = \langle \hat{\textbf{S}}_{\textbf{d}} \cdot \hat{\textbf{S}}_{\textbf{0}} \rangle$ in the undoped system,
\begin{equation}
V_{1}(\ell_\Sigma) = 2 J \l c_{\textbf{e}_x + \textbf{e}_y} - c_{\textbf{e}_x} \r \ell_\Sigma + \delta_{\ell_\Sigma,0} J \l c_{\textbf{e}_x} - c_{2\textbf{e}_x} \r,
\end{equation}
see Sec. VI A in Ref.~\cite{Grusdt2018}. 

Here $\textbf{e}_{x,y}$ denote unit vectors in $x$ and $y$ directions and we assumed for simplicity that the system is isotropic along $x$ and $y$. Note that the spin correlators $c_{\textbf{d}}$ defined above are related to the spin correlations $C(\textbf{0},\textbf{d})$ measured along $z$-direction in the experiment by
\begin{equation}
c_{\textbf{d}} = \langle \hat{\textbf{S}}_{\textbf{0}} \cdot \hat{\textbf{S}}_{\textbf{d}} \rangle = 3 \langle \hat{S}^z_{\textbf{0}} \cdot \hat{S}^z_{\textbf{d}} \rangle = \frac{3}{4} C(\textbf{0},\textbf{d}),
\end{equation}
assuming an $SU(2)$ invariant statistical ensemble as realized in the experiment.

\emph{Relation to physical states.--}
To relate the basis states $|{\textbf{j}^s,\Sigma,\alpha}\rangle$ in the effective theory to physical states, we first identify the trivial string configuration $\Sigma=0$, with string length $0$, with the locally doped spin system:
\begin{equation}
|{\textbf{j}^s,0,\alpha}\rangle \equiv \hat{c}^\dagger_{\textbf{j}^s,\alpha} |{\Psi_0}\rangle = \hat{d}^\dagger_{\textbf{j}^s}  \hat{b}^\dagger_{\textbf{j}^s,\alpha} |{\Psi_0}\rangle.
\end{equation}
Here $\hat{c}^\dagger_{\textbf{j}^s,\alpha}$ denotes a microscopic fermion in the underlying Hubbard model, which can be represented by the combination $\hat{d}^\dagger_{\textbf{j}^s}  \hat{b}^\dagger_{\textbf{j}^s,\alpha}$ of a doublon and a Schwinger boson, and $|{\Psi_0}\rangle$ is the ground state of the undoped spin system. Similarly, at finite temperature we identify the projector $|{\textbf{j}^s,0,\alpha}\rangle \langle{\textbf{j}^s,0,\alpha}|$ with the density matrix $\hat{c}^\dagger_{\textbf{j}^s,\alpha} \hat{\rho}_0 \hat{c}_{\textbf{j}^s,\alpha}$, where $\hat{\rho}_0$ is the thermal state of the undoped spin system.

Non-trivial string states $|{\textbf{j}^s,\Sigma,\alpha}\rangle$ are subsequently obtained by applying tunneling terms $\sum_\beta \hat{c}^\dagger_{\textbf{j}_2,\beta} \hat{c}_{\textbf{j}_1,\beta}$ to move the doublon along a given string $\Sigma$ starting from the state $|{\textbf{j}^s,0,\alpha}\rangle$. Here the string is equal to the doublon trajectory, up to self-retracing paths. Because this doublon motion along the strings $\Sigma$ modifies the positions of the spin background, but not their quantum states, they are refereed to as \emph{geometric strings}, see Refs.~\cite{Grusdt2018,Grusdt2018mixD}. 

For a given string state $\Sigma$, this geometric string construction allows us to relate the correlations between two spins at sites $\textbf{j}_1$ and $\textbf{j}_2$ to spin correlations $c_{\textbf{d}}$ in the original undoped spin system, where $\textbf{d} = \tilde{\textbf{j}}_1 - \tilde{\textbf{j}}_2$ is distance between the two spins at sites $\tilde{\textbf{j}}_1$ and $\tilde{\textbf{j}}_2$ \emph{before} the doublon was moved. By sampling the spinon-chargon states according to their thermal distribution and using the known spin correlations in the undoped spin system,
\begin{flalign}
c_{\textbf{e}_x} &= -0.132, \quad c_{\textbf{e}_x+\textbf{e}_y} = 0.029, \label{eqDefCdUndoped1} \\
 c_{2 \textbf{e}_x} &= 0.015, \quad c_{2 \textbf{e}_x + \textbf{e}_y} = -0.012,
\label{eqDefCdUndoped2}
\end{flalign}
thus allows us to calculate the spin environment of the chargon, i.e., equivalently, the doublon. Spin correlators at larger distances than in Eqs.~\eqref{eqDefCdUndoped1}, \eqref{eqDefCdUndoped2} were experimentally measured to be consistent with zero within error bars and we ignore them in our theoretical analysis. 

Note that, while the harmonic trapping potential is included in the effective string potential, the spin system is much less sensitive to the trap and we assume a homogeneous spin background. As in our comparison to exact numerics, the spin correlations $c_{\textbf{d}}$ in Eqs.~\eqref{eqDefCdUndoped1}, \eqref{eqDefCdUndoped2} are determined by post-selecting on singly occupied sites, as required for our model based on the $t-J$ Hamiltonian.

The spinon-chargon description described above is based on the so-called frozen spin approximation: In the string states $|{\textbf{j}^s,\Sigma,\alpha}\rangle$, the doublon only moves around the surrounding spins. Except for the spinon dynamics at the other end of the string, no back-action of the spin environment is included. This approximation can be justified at strong couplings, $t \gg J$. In this regime, relevant to our experiments performed at $t= 3.5 J$, we can average over fast fluctuations between a large number of string configurations, leading to an overall small effect on the quantum states of the surrounding spins \cite{Grusdt2018}.

\emph{Thermal spinon-chargon states.--}
We solve the spinon-chargon problem in Eq.~\eqref{eqDefHspinonChargon} at strong couplings, $t \gg J$, by a Born-Oppenheimer approximation. First we consider a fixed spinon position $\textbf{j}^s$ and solve for the fluctuating string. This is done using exact numerical diagonalization, and by truncating the Hilbert space at a maximum string length of $\ell_{\rm max}=7$. This yields a thermal state 
\begin{equation}
\hat{\rho}_\Sigma(\textbf{j}^s) = \exp \left[ - \beta ~ \langle{\textbf{j}^s}| \hat{\mathcal{H}}_\Sigma + \hat{\mathcal{H}}_{\rm int} |{\textbf{j}^s}\rangle \right] / ~ Z(\textbf{j}^s),
\end{equation}
where $\beta = 1/ k_{\rm B} T$ and $Z(\textbf{j}^s)$ guarantees proper normalization, $\text{tr} ~ \hat{\rho}_\Sigma(\textbf{j}^s) = 1$.

Next we solve the resulting spinon problem, with the effective Hamiltonian
\begin{equation}
\hat{\mathcal{H}}_s^{\rm eff} = \hat{\mathcal{H}}_s + \sum_{\textbf{j}^s} |{\textbf{j}^s}\rangle \langle{\textbf{j}^s}|   ~ \text{tr} \left[ \hat{\mathcal{H}}\hat{\rho}_\Sigma(\textbf{j}^s) \l  \hat{\mathcal{H}}_\Sigma + \hat{\mathcal{H}}_{\rm int} \r  \right].
\label{eqHeffSpnon}
\end{equation}
We restrict the allowed spinon positions to a central Mott insulating region, measuring $2, 4, 5, 5, 4$ and $2$ sites along $x$-direction at increasing $y$ coordinates. The spinon dispersion $\omega_s(\textbf{k})$ is assumed to be of the form,
\begin{multline}
\omega_s{\textbf{k}} = J \bigl[ \lambda_1 \l \cos(2 k_x) + \cos(2 k_y) \r +\\
 2 \lambda_2 \l \cos(k_x+k_y) + \cos(k_x-k_y) \r \bigr]
\end{multline}
which fits well to exact numerical Monte Carlo results for the magnetic polaron dispersion by Brunner et al. \cite{Brunner2000}. As fit parameters we use $\lambda_1 = 0.28$ and $\lambda_2 = 0.16$. We have checked that our results for the spin correlations around the doublon do not depend strongly on the spinon dispersion at the experimentally relevant temperatures $T = 1.4 J$. The single-particle problem in Eq.~\eqref{eqHeffSpnon} can now be easily solved, and we obtain the thermal state
\begin{equation}
\hat{\mathcal{H}}\hat{\rho}_s = \exp \left[ - \beta \hat{\mathcal{H}}_s^{\rm eff}  \right] / ~ Z_s,
\end{equation}
from which we sample spinon positions.

%
\bibliography{bibliography}

\end{document}